\documentclass[sigconf]{aamas} 

\usepackage{balance} 
\usepackage{algorithm}
\usepackage{algorithmic}
\usepackage{multirow}

\setcopyright{ifaamas}
\acmConference[AAMAS '24]{Proc.\@ of the 23rd International Conference
on Autonomous Agents and Multiagent Systems (AAMAS 2024)}{May 6 -- 10, 2024}
{Auckland, New Zealand}{N.~Alechina, V.~Dignum, M.~Dastani, J.S.~Sichman (eds.)}
\copyrightyear{2024}
\acmYear{2024}
\acmDOI{}
\acmPrice{}
\acmISBN{}

\acmSubmissionID{<<65>>}

\title[AAMAS-2024 Formatting Instructions]{Developing A Multi-Agent and Self-Adaptive Framework with Deep Reinforcement Learning\\ for Dynamic Portfolio Risk Management}

\author{Zhenglong Li}
\affiliation{
  \institution{The University of Hong Kong}
  \city{Hong Kong}
  \country{China}}
\email{lzlong@hku.hk}

\author{Vincent Tam}
\affiliation{
  \institution{The University of Hong Kong}
  \city{Hong Kong}
  \country{China}}
\email{vtam@eee.hku.hk}

\author{Kwan L. Yeung}
\affiliation{
  \institution{The University of Hong Kong}
  \city{Hong Kong}
  \country{China}}
\email{kyeung@eee.hku.hk}

\begin{abstract}
Deep or reinforcement learning (RL) approaches have been adapted as reactive agents to quickly learn and respond with new investment strategies for portfolio management under the highly turbulent financial market environments in recent years. In many cases, due to the very complex correlations among various financial sectors, and the fluctuating trends in different financial markets, a deep or reinforcement learning based agent can be biased in maximising the total returns of the newly formulated investment portfolio while neglecting its potential risks under the turmoil of various market conditions in the global or regional sectors. Accordingly, a multi-agent and self-adaptive framework namely the MASA is proposed in which a sophisticated multi-agent reinforcement learning (RL) approach is adopted through two cooperating and reactive agents to carefully and dynamically balance the trade-off between the overall portfolio returns and their potential risks. Besides, a very flexible and proactive agent as the market observer is integrated into the MASA framework to provide some additional information on the estimated market trends as valuable feedbacks for multi-agent RL approach to quickly adapt to the ever-changing market conditions. The obtained empirical results clearly reveal the potential strengths of our proposed MASA framework based on the multi-agent RL approach against many well-known RL-based approaches on the challenging data sets of the CSI 300, Dow Jones Industrial Average and S\&P 500 indexes over the past 10 years. More importantly, our proposed MASA framework shed lights on many possible directions for future investigation.
\end{abstract}

\keywords{Deep Reinforcement Learning; Multi-Agent; Risk Management; Self-Adaptive; Portfolio Optimisation}

\newcommand{\BibTeX}{\rm B\kern-.05em{\sc i\kern-.025em b}\kern-.08em\TeX}

\makeatletter
\gdef\@copyrightpermission{
	\begin{minipage}{0.3\columnwidth}
		\href{https://creativecommons.org/licenses/by/4.0/}{\includegraphics[width=0.90\textwidth]{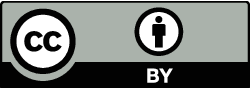}}
	\end{minipage}\hfill
	\begin{minipage}{0.7\columnwidth}
		\href{https://creativecommons.org/licenses/by/4.0/}{This work is licensed under a Creative Commons Attribution International 4.0 License.}
	\end{minipage}
	\vspace{5pt}
}
\makeatother

\begin{document}


\pagestyle{fancy}
\fancyhead{}


\maketitle 


\section{Introduction}

Computational Finance (CF)~\cite{gunjan2023brief,maglaras2022deep,tsang2023ai} is a very active research area involving the studies of computational approaches to tackle many different challenging and practical problems in Finance. Conventionally, algorithmic methods had been employed to simulate various investment strategies and their plausible results in the financial markets. Recently, many researchers have tried to explore the potential uses of machine learning approaches~\cite{nabipour2020predicting} including the support vector machines~\cite{sivaram2020optimal}, deep learning (DL) or reinforcement (RL) learning approaches~\cite{jiang2017deep,wang2021hierarchical,xu2021relation,yang2022smart} in a diversity of real-world applications~\cite{duan2022factorvae,liang2021adaptive,liu2020adaptive,zhang2020cost} in CF. 
Among these applications, DL or RL approaches such as the Twin Delayed DDPG (TD3) algorithm~\cite{fujimoto2018addressing} for dynamic environments with continuous action spaces have been adapted as reactive agents~\cite{cuschieri2021td3} to quickly learn and respond with new investment strategies for portfolio management under the highly turbulent financial market environments in recent years. In many cases, due to the very complex correlations among various financial sectors, and the fluctuating trends in different financial markets, a deep or reinforcement learning based agent can be mainly focused on maximising the total returns of the newly formulated investment portfolio while ignoring the potential risks of the new investment portfolio under the turmoil of various market conditions such as the unpredictable and sudden changes of the market trends frequently occurring in the global or regional sectors of financial markets, especially due to the COVID-19 pandemic, 
natural disasters brought by extreme weathers, and local conflicts across different regions, etc. 

To overcome the above pitfall, a {\bf m}ulti-{\bf a}gent and {\bf s}elf-{\bf a}daptive framework namely the {\it MASA\/} is proposed in this work in which two cooperating and reactive agents are utilised to 
implement a radically new multi-agent RL scheme so as to
carefully and dynamically balance the trade-off between the overall returns of the newly revised portfolio 
and their potential risks especially when the concerned financial markets are highly turbulent.
The first cooperating agent is based on the TD3 algorithm targeted to optimise the overall returns of the current investment portfolio while the second intelligent agent is based on a complete constraint solver, or possibly any efficient local optimisers such as the evolutionary algorithms~\cite{storn1997differential} or particle swarm optimisation (PSO) methods~\cite{kennedy1995particle}, trying to adjust the current investment portfolio in order to minimise its potential risks after considering the estimated market trend as provided by another adaptive agent as the market observer in the proposed MASA framework.
Clearly, the multi-agent RL scheme of the proposed MASA framework may help to produce more balanced 
investment portfolios
in terms of both portfolio returns and potential risks with the clear division of works between the two cooperating agents to continuously learn and adapt from the underlying financial market environment. 
It is worth noting that multi-agent RL-based frameworks have been actually considered in some previous research studies. 
For instance, a TD3-based multi-agent deep reinforcement learning (DRL) approach~\cite{ZHANG2020206} was investigated 
in a previous work to improve the function approximation error and complex mission adaptability through applications to the mixed cooperation-competition environment in a general perspective. 
Yet instead of relying on the complex and dual-centered Q-network to reduce the bias of function estimation 
as in the previous work, 
our proposal has uniquely focused on using the TD3-based agent to firstly optimise on the overall returns
of the newly revised portfolio with some possibly under-estimated bias/error in its potential risks to be quickly rectified 
by the second solver-based agent using a loosely-coupled and pipelining computational model to tackle this 
specific and challenging problem of dynamic portfolio risk management in the real-world applications of CF.  
It should be noted that by adopting the loosely-coupled and pipelining computational model, the proposed MASA framework will become more resilient and reliable since the overall framework will continue to work even when any particular agent fails.  
Moreover, 
to make the proposed MASA framework more adaptive to the extremely volatile environments of financial markets,
the market observer as a very flexible and proactive agent to continuously provide 
the estimated market trends as valuable feedbacks 
for the other two cooperating agents to quickly adapt to the ever-changing market conditions. 
Undoubtedly, this simply highlights another key difference of our proposal on the multi-agent RL scheme 
when compared to those multi-agent RL-based frameworks examined in the previous studies.
Furthermore, when the market observer agent is implemented as a deep neural network such as 
the multi-layer perceptron (MLP)~\cite{wang2021dcn} model, the resulting MASA framework can be extended as 
a DRL approach for dynamic portfolio management in CF.

To demonstrate the effectiveness of our proposal, a prototype of the proposed MASA framework is implemented in Python and tested on a GPU server installed with the Nvidia RTX 3090 GPU card.
The attained empirical results demonstrate the potential strengths of our proposed MASA framework based on the multi-agent RL approach against many well-known RL-based approaches on the challenging data set of the CSI 300, 
Dow Jones Industrial Average (DJIA) and S\&P 500 indexes over the past 10 years. More importantly, our proposed MASA framework shed lights on many possible directions including the exploration of utilising different meta-heuristic based optimisers such as the PSO for the solver-based agent, various machine learning approaches for the market observer agent, 
or the potential applications of the proposed MASA framework for various resource allocation, planning or disaster recovery problems in which  the risk management is very critical 
for our future investigation.



\section{The Preliminaries} \label{sec:prelim}

\subsection{Reinforcement Learning} 
As one of the active research areas in machine learning~\cite{goodell2021artificial}, RL~\cite{kolm2020modern} is mainly focused on how intelligent agents make rational decisions on actions based on specific observations in a possibly unexplored environment in order to maximise the cumulative rewards of the performed actions with respect to the underlying environment. The key focus of RL approaches is to strive for a {\it balance\/} between the {\em exploration\/} of the unknown environment and the {\em exploitation\/} of the current knowledge gained through the iterative learning process.
The underlying environment is usually stated in the form of a partially observable Markov decision process (POMDP)~\cite{puterman1990markov} in which dynamic programming techniques~\cite{markowitz2000mean} can be employed to solve the involved POMDP. Yet the RL approaches are targeted to handle large POMDPs where exact methods like the dynamic programming techniques may fail since the RL approaches do not need to assume any prior knowledge of the involved POMDP to represent the underlying environment. Clearly, the RL approaches are very suitable to explore the uncharted and also unpredictable environments of various financial markets when solving a diversity of real-world problems in CF.

In recent years, the RL approaches have attained remarkable successes for portfolio optimisation in which RL-based investment strategies have demonstrated adaptive and fast learning abilities to adjust the portfolios for maximising the overall returns after a targeted trading period. 
Among the numerous RL approaches, a successful example is the TD3 algorithm~\cite{fujimoto2018addressing} as a model-free, online, off-policy reinforcement learning method. 
Generally speaking, a TD3 agent is an actor-critic reinforcement learning agent that is aimed to look for an optimal policy to maximise the expected cumulative long-term reward. 
For portfolio optimisation, the expected cumulative long-term reward of the TD3 agent can be straightforwardly formulated as the expected overall returns of the concerned portfolio after a specific trading period.
Yet with the highly volatile financial market conditions, 
it can be difficult for most RL approaches to strive for a good balance between the intrinsically conflicting objectives of 
maximising returns and also minimising the risks of portfolios  
over a specific trading period. In most cases with turbulent market conditions,
increasing the portfolio returns will likely increase the potential risks that may 
possibly lead to great and sudden losses of the investment portfolio in an extremely short period of time
due to some unexpected crises. 

\subsection{Multi-Agent Systems}
Multi-agent systems (MAS)~\cite{Kampouridis_2022}  is a core and very active research area of 
artificial intelligence~\cite{zheng2019finbrain}  in which 
many different perspectives and methodologies including the neural networks~\cite{hafezi2015bat} or evolutionary algorithms~\cite{drezewski2017agent}  have been adopted and contributed to the latest development of MAS. 
In many real-world applications such as various challenging problems in CF, 
multiple intelligent agents may try to optimise their own returns and/or other objective(s), 
that may unavoidably collide with the interests of other investors with the same objective(s). 

 Besides, there are other research studies~\cite{Calliess_2021, belcak_2022} describing how MAS may facilitate the simulations in research studies of 
CF and Computational Economics.
To more precisely model from the perspective of MAS,
each agent in the multi-agent market simulation environment may find it difficult to 
learn a static investment strategy due to the fluctuating market dynamics.
Thus, the involved agents may need to deploy intelligent algorithms capable of learning to compete well with adaptive mechanisms in the adversarial market environments.
In addition, 
studying intelligent trading through such simulations from a multi-agent perspective can lead to many exciting research directions with possible findings of relevance to policy makers and investors. An example is the market simulator (MAXE) \cite{belcak_2022} 
to examine different types of agent behaviour, market rules and anomalies on market dynamics through the simulation of large-scale MAS.

It is worth noting that there are some previous research studies investigating 
the potential uses of RL-based algorithms in MAS for many applications. 
For instance, a TD3-based multi-agent DRL approach~\cite{ZHANG2020206} was examined 
in a previous work to improve the function approximation error and complex mission adaptability through applications to the mixed cooperation-competition environment 
from a general perspective. 
Essentially, the TD3-based DRL approach makes use of 
the complex and dual-centered Q-network to reduce the bias of function estimation.
On the other hand, 
our proposal of the MASA framework 
has focused on using the TD3-based agent to firstly optimise on the overall returns
of the newly revised portfolio while leaving the potential risks as the possibly under-estimated error 
to be effectively handled by the second solver-based agent  
with a loosely-coupled and pipelining mechanism for 
dynamic portfolio risk management in CF.  
Through adopting the loosely-coupled and pipelining computational model, the proposed MASA framework will become a dependable MAS with high availability and reliability 
since the resulting MAS will continue to work even when any specific agent fails.  

\subsection{Portfolio Optimisation in Computational Finance}
Portfolio optimisation is a very challenging multi-objective optimisation problem in CF where 
the uncharted and highly volatile financial market environments can be difficult for many intelligent algorithms or 
well-known mathematical programming~\cite{sawik2012bi} approaches to tackle.
Conventionally,  many investors and researchers utilised specific financial indicators such as the moving averages~\cite{raudys2013moving} or the relative strength index~\cite{alfaro2010financial}, together with the heuristic or machine learning approaches   
including the follow-the-winner, follow-the-loser, pattern-matching or meta-learning algorithms~\cite{li2014online} 
to try to capture the momentum of price changes. 
Recently, there have been many interesting research studies trying to apply DL or RL techniques to explore the 
turbulent and uncharted financial market environments. For instance, 
\cite{liang2021adaptive,ye2020reinforcement} 
consider the news data as an additional information for portfolio management 
while \cite{xu2021relation,jiang2017deep} utilise specific modules as intelligent agents 
to carefully deal with the assets information and then capture the correlations among the involved assets. 




To facilitate our subsequence discussion, some essential concepts including the 
portfolio value, both the short-term and long-term risks of a portfolio, etc. 
related to portfolio management in CF are given as below. 

\begin{definition} \label{def:portfolio}
(Portfolio Value) The total value of a portfolio at time $t$  is
\begin{equation}
    C_t = \sum_{i=1}^{N} a_{t,i} \times p^{c}_{t,i},
\end{equation}
\end{definition}
\noindent where $N$ is the number of assets in a portfolio, $a_{t,i}$ is the weight of $i^{\text{th}}$ asset, and $p^{c}_{t,i}$ is the close price of $i^{\text{th}}$ asset at time $t$. 

Accordingly, each investment portfolio is constrained as below.
 \begin{equation}
\forall a_{t,i}\in \mathbf{A}_t:\quad a_{t,i}\geq 0 ,\sum_{i=1}^N a_{t,i}=1,
\end{equation}
\noindent where $\mathbf{A}_t \in \mathbf{A}$ is the weight vector $\mathbf{A}$ at time $t$. 
Clearly, the summation of all the allocation weights $a_{t,i}$ for the total $N$ assets of a complete portfolio should be $1$.

Based on the well-known Markowitz model \cite{markowitz2000mean},
both the short-term and long-term risks of a portfolio over a specific trading period can be defined 
in terms of  the corresponding covariance-weighted risk and the volatility of strategies as follows.
\begin{definition}
\label{def:shorttermrisk}
 (Short-term Portfolio Risk) The short-term portfolio risk $ \sigma_{p,t}$ at time $t$ is defined as below.
 \begin{equation}
    \begin{aligned}
    \sigma_{p,t} & = \sigma_\beta + \sigma_{\alpha,t}\\
    \sigma_{\alpha,t} & = \sqrt{\mathbf{A}^T_{t}\Sigma_k \mathbf{A}_{t}} = \Vert \Sigma_k \mathbf{A}_{t} \Vert_2,
    \end{aligned}
\end{equation}
\end{definition}
\noindent
\noindent where $\sigma_{\alpha,t}$ is the trading strategy risk, $\sigma_\beta$ is the market risk and $\mathbf{A}_{t}\in \mathcal{R}^{N \times 1}$ is the matrix of weights. The covariance matrix $\Sigma_k\in\mathcal{R}^{N\times N}$ between any two assets can be calculated by the rate of daily returns of assets in the past $k$ days. 

\begin{definition} \label{def:longtermrisk}
(Long-term Portfolio Risk) The long-term portfolio risk $Vol_p$ is defined as the strategy volatility that is the sampled variance of the daily return rates $r_{p,t}$ of a trading strategy over the whole trading period. $\overline{r}_{p,t}$ is the average daily return rate.

 \begin{equation}
    \begin{aligned}
    Vol_{p}=\sqrt{\frac{252}{T-1} \sum_{t=1}^{T}\left(r_{p,t}-\overline{r}_{p,t}\right)^{2}}.
    \end{aligned}
\end{equation}

\end{definition}

Besides, the following gives a formal definition of the Sharpe ratio as 
one of the most widely adopted performance measures on the 
risk-adjusted relative returns of a portfolio.
\begin{definition}\label{def:sharperatio}
(Sharpe Ratio) The Sharpe Ratio (SR) is a performance indicator for evaluating a portfolio in terms of the total annualized returns $R_p$, risk-free rate $r_f$ and annualized long-term portfolio risk $Vol_p$. 
 \begin{equation}
\text{SR} = \frac{R_{p}-r_f}{Vol_p}.
\end{equation}
\end{definition}

More importantly, it should be noted that the portfolio optimisation problem in CF is used in this work to demonstrate the feasibility of our proposed MASA framework for risk management under the highly volatile and unknown environments.
In the future investigation, it would be interesting to explore how 
the multi-agent RL-based approach of the proposed MASA framework can be adapted to various planning or resource allocation
problems under certain hostile and unknown environments such as those for disaster recovery or emergency management.


\section{The Proposed Multi-Agent and Self-Adaptive Framework} \label{sec:MASA} 

To overcome the pitfall of the RL-based approaches to bias on optimising the investment returns, 
a {\bf m}ulti-{\bf a}gent and {\bf s}elf-{\bf a}daptive framework namely the {\it MASA\/} is proposed in this work 
in which two cooperating and reactive agents, namely the RL-based and solver-based agents, 
are utilised to implement a radically new multi-agent RL scheme in order to
dynamically balance the trade-off between the overall returns of the newly revised portfolio 
and potential risks especially when the financial markets are highly turbulent.

\begin{figure}[h] 
    \centering
    \includegraphics[width=1.0\linewidth,height=1.4in]{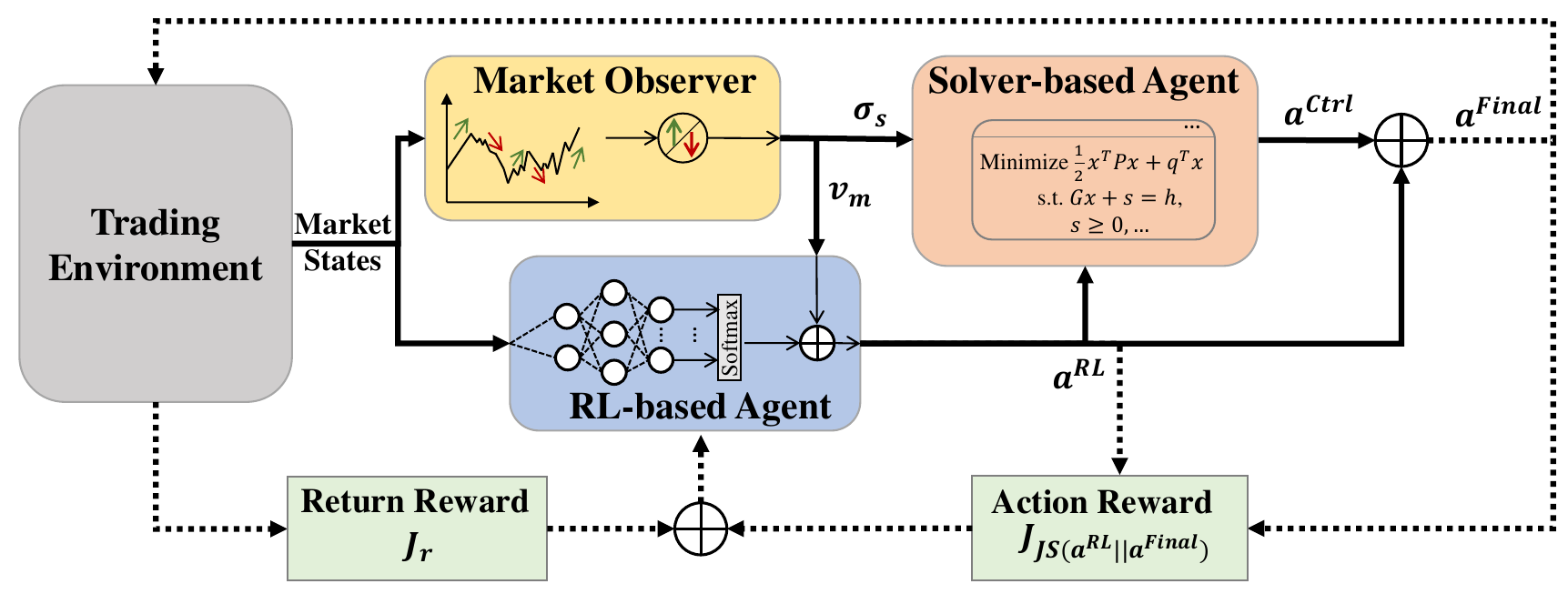}
    \caption{The System Architecture of the Proposed MASA Framework}
    \label{fig:MASA_sys_diag}
\end{figure}
Figure \ref{fig:MASA_sys_diag} reviews the overall system architecture of the proposed MASA framework
in which the RL-based agent is based on the TD3 algorithm to optimise the overall returns of the current investment portfolio while the solver-based agent is based on a complete constraint solver, or possibly any efficient local optimisers such as the evolutionary algorithms~\cite{storn1997differential} or PSO methods~\cite{kennedy1995particle}, that works to further adapt the investment portfolio returned by the RL-based agent so as to minimise its potential risks after considering the estimated market trend as provided by the market observer of the proposed MASA framework.
In essence, 
through the clear division of works between both RL-based and solver-based agents to 
continuously learn and adapt from the underlying financial market environment with the support by market observer agent,
the multi-agent RL scheme of the proposed MASA framework may help to attain 
more balanced investment portfolios
in terms of both portfolio returns and potential risks 
when compared to those portfolios returned by the RL-based approaches. 
It is worth noting that the proposed MASA framework 
adopts a loosely-coupled and pipelining computational model among the three cooperating and intelligent agents,
thus making the overall multi-agent RL-based approach
more resilient and reliable since the overall framework will continue to work in the worst case of 
any individual agent being failed.  

In addition,
to make the proposed MASA framework more adaptive to the extremely volatile environments of financial markets,
the market observer agent will continuously provide the estimated market trends as valuable feedbacks 
for both RL-based and solver-based agents to quickly adapt to the ever-changing market conditions. 
Furthermore, when the market observer agent is implemented as a deep neural network such as 
the MLP~\cite{wang2021dcn} model, the resulting MASA framework can be extended as 
a multi-agent DRL approach for dynamic portfolio management in CF.
The empirical evaluation results of the market observer agent implemented as 
an algorithmic approach \cite{tsang2017profiling}, the MLP and another deep learning models are carefully analysed in Section \ref{sec:eval}.

\begin{algorithm}[tb]
    \caption{The Training Procedure of the MASA Framework}
    \label{algo:MASA} 
    \begin{algorithmic}[1]
        \STATE \textbf{Input}: $T$ as the total number of trading days, $MaxEpisode$ as the maximum number of episodes, the settings of RL-based agent, and the selected market observer agent. 
        \STATE \textbf{Output}: The revised RL policy $\pi^{*}$ and (possibly) updated market observer agent. 
        \STATE Initialise the RL policy $\pi_0$, the market observer agent and memory tuple $\hat{D}$ and $\hat{M}$.
        \FOR{$k=1$ to $MaxEpisode$}
        \STATE Reset the trading environment and set the initial action $a_0^{Final}$ and $a_0^{RL}$.
        \FOR{$t=1$ to $T$}
        \STATE Observe the current market state $o_t$
        \STATE Calculate the reward $r_{t-1}$ by $a^{Final}_{t-1}$
        \STATE Store tuple ($o_{t-1}$, $a^{Final}_{t-1}$, $a^{RL}_{t-1}$, $o_t$, $r_{t-1}$) in $\hat{D}$
        \STATE Store tuple ($o_{t-1}$, $o_{t}$, $\sigma_{s,t-1}$, $v_{m,t-1}$) in $\hat{M}$
        \STATE Invoke the market observer agent to compute the suggested risk boundary $\sigma_{s,t}$ and market vector $v_{m,t}$ as the additional feedback for updating both the RL-based and solver-based agents
        \STATE Invoke the RL-based agent $\pi_{t}$ to generate the current action $a_t^{RL}$ as portfolio weights
        \STATE Invoke the solver-based agent to generate the adjusted action $a_t^{Ctrl}$
        \STATE Adjust the current portfolio by $a_t^{Final}=a_t^{RL}+a_t^{Ctrl}$
        \STATE Execute the portfolio order with $a_{t}^{Final}$
        \IF {the RL policy update condition is triggered}
        \STATE Update the RL policy $\pi$ by learning the historical trading data from $\hat{D}$
        \ENDIF
        \IF {the predefined update condition of the market observer agent is triggered}
        \STATE Update the market observer agent by learning the historical profile $\hat{M}$
        \ENDIF
        \ENDFOR
        \ENDFOR
        \STATE \textbf{return} the best RL policy $\pi^*$ and the possibly updated market observer agent
    \end{algorithmic}
\end{algorithm}
The pseudo-code of the training procedure of the proposed MASA framework is shown in Algorithm \ref{algo:MASA} 
to illustrate how the 3 cooperating agents are working with each other to adaptively achieve the conflicting objectives 
of optimising returns and minimising risks in response to the possibly highly turbulent financial market conditions.
Firstly, before the iterative training process is started, all the relevant information including the RL policy, 
the market state information stored in the market observer agent, etc. are initialised.
During the training process, the current market state information $o_t$ such as the most recent downward or upward trend 
of the underlying financial market over the past few trading days will be collected as the basic information
for the subsequent computation of the market observer agent.
 Besides, the reward of the previously executed action $a^{Final}_{t-1}$ will be computed as the feedback for the RL-based algorithm to revise its RL policy.
The market observer agent will then be invoked to compute the
suggested risk boundary $\sigma_{s,t}$ and market vector $v_{m,t}$ as 
some additional feedback for updating both the RL-based and solver-based agents on the latest market conditions.
As aforementioned, to maintain the flexibility and self-adaptivity of the proposed MASA framework,
there can be various approaches including 
the algorithmic approach such as the directional changes \cite{tsang2017profiling,bakhach2016forecasting}, 
deep neural networks such as the MLP or other DL approaches \cite{hochreiter1997long,wang2021deeptrader} 
that can be considered in more detail in Section \ref{sec:eval}.
More importantly, it should be noted that both the RL-based and solver-based agents are already secured with the current market information as the most valuable feedback 
obtained from the existing trading environment as shown in Figure \ref{fig:MASA_sys_diag}.
The provision of the suggested market condition information by the market observer agent 
are used solely as an additional information to quickly adapt and enhance the performance of 
both the RL-based and solver-based agents especially when the latest market conditions are highly volatile.
In the worst cases when the suggested market condition produced by the market observer agent can be incorrect 
as 'noises' to mislead the search of both RL-based and solver-based agents for possibly biased actions in specific trading days,
the {\em self-adaptive nature\/} of the reward mechanism of the RL-based agent to adapt from the underlying trading environment in the subsequent iterations of training, and also the {\em auto-corrective learning capability\/} of the intelligent market observer 
algorithm will help to ensure that such misleading noises can be effectively and quickly fixed over a longer period of trading to gain more valuable domain knowledge and insights about the underlying market conditions through updating the 
learning history profile $\hat{M}$ of the market observer agent.
Interestingly, as observed from the empirical evaluation results obtained in Section \ref{sec:eval}, there can be fairly impressive enhancements in the ultimate performance of both RL-based and solver-based agents even when some relatively simple algorithmic approach based on the directional changes is used to implement the market observer agent in the proposed MASA framework on the challenging data sets of the CSI 300, DJIA  and S\&P 500 indexes over the past 10 years. 
Clearly, for a deeper understanding of the ultimate impacts of the suggested information by the market observer agent to 
the other two intelligent agents, the proposed MASA framework should be applied to 
more challenging data sets in CF or other application domains for more in-depth and thorough analyses in the future research studies. After the market observer agent is invoked, 
the RL-based agent will be triggered to generate the current action $a_t^{RL}$ as portfolio weights that can be further revised
by the subsequent solver-based agent after considering its own risk management strategy and the suggested market condition 
provided by the market observer agent. All in all, through adopting this loosely-coupled and pipelining computational model,
the resulting MASA framework will continue to work as a dependable MAS even when any individual agent fails.

\begin{figure}[ht]
  \centering
  \includegraphics[width=0.9\linewidth]{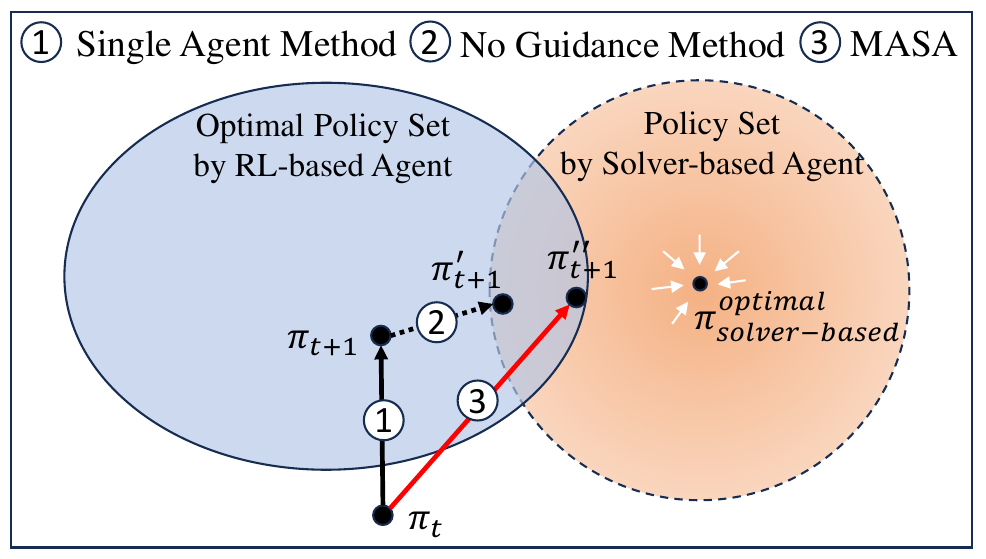}
  \caption{An Illustration of the Guiding Mechanism of the MASA Framework to Gradually Enhance the Constructed Policies of the RL-Based Agent}
  \label{fig:policycomp}
  \Description{The Changes of Portfolio Values of the Policies by the RL-Based Agent on the DJIA Index}
\end{figure}
Figure \ref{fig:policycomp} demonstrates the strengths of the reward-based guiding mechanism 
adopted by the proposed MASA framework to gradually
enhance the various policies constructed by the single-agent RL-based approach, the proposed MASA framework without 
the reward-based guiding mechanism, and the proposed MASA framework utilising the reward-based guiding mechanism.
The single-agent RL-based approach can update the policy $\pi_{t+1}$ into the relatively more optimal set (i.e., the blue shaded area) by maximising the total returns of portfolios yet it may possibly neglect the potential risks. 
On the other hand, the red shaded area of  Figure \ref{fig:policycomp}  
represents the policy set as recommended by the solver-based agent to minimise the potential risks for portfolio management. 
When working independently, each of the agents cannot combine the best advantages to achieve a more optimal portfolio 
for both objectives on the overall returns and potential risks.
Besides, as shown in Figure \ref{fig:policycomp},
when all the 3 proposed agents working together without any intelligent guiding mechanism such as the reward-based guidance,
the resulting framework can be easily stuck in specific local minima. 
Accordingly, through the reward-based guiding mechanism as adopted by the 
MASA framework to carefully respond to the ever-changing environment,
both the RL-based and solver-based agents can iteratively enhance the current investment portfolio 
with respect to both objectives of the overall returns and potential risks after considering the valuable feedback from the third 
market observer agent. At the same time, the reward-based guiding mechanism of the MASA framework utilises 
an entropy-based divergence measure such as the Jensen–Shannon divergence (JSD)~\cite{MENENDEZ1997307, lin1991}
for promoting the diversity of the generated action sets 
as an intelligent and self-adaptive strategy to cater for the highly volatile environments of various financial markets.

The augmented reward function for the RL-based agent is depicted as follows.
\begin{equation}
    \label{eqn:augreward}
     J\left ( \theta  \right ) = \lambda_1 J_{r}\left ( \theta \right ) + \lambda_2  J_{JS}\left ( \theta \right ),
\end{equation}
\noindent where $\lambda_1$ and $\lambda_2$ are the learning rates of the return reward $J_{r}(\theta)$ and the action reward $J_{JS}(\theta)$. To maximise the overall returns of the current investment portfolio, the $J_{r}(\theta)$ can be computed as the sum of the logarithm of returns as stated in Equation~\eqref{eqn:returnreward}.
    \begin{equation}
    	\label{eqn:returnreward}
    		\begin{aligned}
            J_{r}\left ( \theta \right ) & = \frac{1}{T} \log{{C_0}\prod_{t=1}^{T} {r_t}} \\
            &=\frac{1}{T} \left (\log{C_0} + \sum_{t=1}^{T} \log{r_t}\right ),
    		\end{aligned}
    \end{equation}
\noindent where $C_0$ is the initial portfolio value, $T$ is the number of trading days, and $r_t=\frac{C_t}{C_{t-1}}$ is the growth rate of portfolio at $t$. 

The action-guided reward $J_{JS}(\theta)$ to promote the diversity of the action sets as generated by the proposed MASA framework 
is defined as below.
    \begin{equation}
    	\label{eqn:actionreward}
            J_{JS}\left ( \theta \right ) = -\frac{1}{T}\sum_{t=1}^{T}{D_{JS}\left (\textbf{a}^{RL}_t \left |  \right | \textbf{a}^{Final}_t \right )},
    \end{equation}
\noindent where $\textbf{a}_t^{RL}$ is the action generated by the RL-based agent at $t$, $\textbf{a}_t^{Final}$ is the adjusted action after considering the actions as recommended by both the RL-based and solver-based agents, 
and $D_{JS}$ is the JSD to measure the similarity between
$\textbf{a}_t^{RL}$ and $\textbf{a}_t^{Final}$ as two probability distributions of the actions generated by the MASA framework.

\section{An Empirical Evaluation} \label{sec:eval} 
\textbf{Datasets}: To demonstrate the effectiveness of the proposed MASA framework in tackling the real-world portfolio risk management  
with conflicting objectives under the mostly uncharted and highly volatile financial market environments,
a preliminary prototype of the proposed MASA framework is implemented in Python, and evaluated on a GPU server machine installed with the AMD Ryzen 9 3900X 12-Core processor running at 3.8 GHz and two Nvidia RTX 3090 GPU cards. Furthermore, the MASA framework is compared with other methods on three challenging yet representative data sets of CSI 300, Dow Jones Industrial Average (DJIA) and S\&P 500 indexes from September 2013 to August 2023 in which the first five-year data is used to train the model, followed by the subsequent data set of two years to validate the trained model. Lastly, all the validated models of various approaches are evaluated on the data set of the latest three years. The top 10 stocks of each index are selected to construct the investment portfolio in terms of the company capital. 
In addition, all the involved data sets contain 
both upward and downward trends of stock prices, and also various patterns of fluctuation for different market conditions
so as to avoid any possible bias toward a specific approach under the evaluation.

\noindent\textbf{Comparative Methods}: Ten representative methods based on algorithmic or RL approaches are carefully selected to compare against the MASA framework. The Constant Rebalanced Portfolio (CRP) \cite{cover1991universal} is the vanilla strategy of equal weighting. The Exponential Gradient (EG) method \cite{helmbold1998line} is based on the follow-the-winner approach while the Online Moving Average Reversion (OLMAR) \cite{li26line}, Passive Aggressive Mean Reversion (PAMR) \cite{li2012pamr}, and Robust Median Reversion (RMR) \cite{huang2016robust} approaches follow the loser assets during trading. The Correlation-driven Nonparametric Learning Strategy (CORN) \cite{li2011corn} is a heuristic strategy to match historical investment patterns. Moreover, the four latest RL-based portfolio optimisation approaches are considered. Ensemble of Identical Independent Evaluators (EIIE) \cite{jiang2017deep} is based on a convolution-based neural network to extract the features of assets while Portfolio Policy Network (PPN) \cite{zhang2020cost} consists of a recurrent-based and a  convolution-based neural networks to capture the sequential information and correlations between assets. Besides, Relation-Aware Transformer (RAT) \cite{xu2021relation} is a transformer-based model to learn the patterns from price series. Lastly, the TD3 with a profit maximisation strategy (TD3-Profit) \cite{fujimoto2018addressing} as the classical RL approach is included for the comparison.

Besides, to evaluate the profitability and risk management of the concerned approaches, four commonly adopted performance metrics 
 including the Annual Return (AR), Maximum Drawdown (MDD), Sharpe Ratio (SR), and short-term portfolio risk (Risk) are considered. 
 Specifically, the SR is a comprehensive metric to indicate the balance between the portfolio returns and risks 
 as attained by each approach.
 All the reported results are averaged over 10 runs. 


\begin{table*}[ht]
  \centering
  \caption{The Performance of Various Well-Known RL-Based Approaches Against the Proposed MASA Framework on Different Challenging Data Sets of Financial Indexes} \label{tab:eval_indexes}
    \resizebox{\textwidth}{!}{
    \begin{tabular}{c|cccc|cccc|cccc}
    \hline
    Market & \multicolumn{4}{c|}{\textbf{CSI 300}}  & \multicolumn{4}{c|}{\textbf{DJIA}} & \multicolumn{4}{c}{\textbf{S\&P 500}} \\
    \hline
    Models & \textbf{AR$(\%)\uparrow$}     & \textbf{MDD$(\%)\downarrow$}     & \textbf{SR$\uparrow$}   & \textbf{Risk$\downarrow$}  & \textbf{AR$(\%)\uparrow$}     & \textbf{MDD$(\%)\downarrow$}     & \textbf{SR$\uparrow$} & \textbf{Risk$\downarrow$} & \textbf{AR$(\%)\uparrow$}     & \textbf{MDD$(\%)\downarrow$}     & \textbf{SR$\uparrow$} & \textbf{Risk$\downarrow$} \\
    \hline

    CRP        & 7.19   & 33.96 & 0.19  & 0.0131 & 11.44  & 19.66 & 0.58  & 0.0095 & 18.09 & 37.12 & 0.65  & 0.0143 \\
    EG         & 7.19   & 33.94 & 0.19  & 0.0131 & 11.39  & 19.66 & 0.57  & 0.0095 & 18.03 & 36.88 & 0.65  & 0.0142 \\
    OLMAR      & -3.50  & 55.67 & -0.17 & 0.0217 & -13.89 & 59.89 & -0.54 & 0.0179 & \textbf{29.34} & 68.52 & 0.55  & 0.0275 \\
    PAMR       & -14.24 & 49.32 & -0.47 & 0.0223 & -37.72 & 81.72 & -1.35 & 0.0174 & 3.63  & 59.08 & 0.05  & 0.0263 \\
    CORN       & -2.06  & 59.28 & -0.13 & 0.0219 & 1.62   & 41.76 & 0.00  & 0.0122 & -6.90 & 62.97 & -0.21 & 0.0200 \\
    RMR        & 5.77   & 41.96 & 0.07  & 0.0215 & -12.98 & 61.23 & -0.51 & 0.0180 & -4.12 & 85.48 & -0.09 & 0.0280 \\
    EIIE       & 6.84   & 31.77 & 0.18  & 0.0122 & 10.81  & 18.24 & 0.58  & 0.0089 & 16.50 & 35.80 & 0.63  & 0.0135 \\
    PPN        & 6.74   & \textbf{31.19} & 0.18  & 0.0119 & 10.50  & 17.95 & 0.57  & 0.0087 & 16.65 & 34.17 & 0.65  & 0.0130 \\
    RAT        & 6.78   & 31.48 & 0.18  & 0.0121 & 10.62  & 18.19 & 0.57  & 0.0088 & 17.03 & 34.92 & 0.65  & 0.0133 \\
    TD3-Profit & 7.18   & 33.97 & 0.19  & 0.0128 & 11.45  & 19.65 & 0.58  & 0.0095 & 18.09 & 37.12 & 0.65  & 0.0143 \\
    \hline
    \textbf{MASA-MLP}  & \textbf{8.87}   & 31.78 & \textbf{0.27}  & \textbf{0.0119} & 13.17  & 19.89 & 0.69  & 0.0088 & 22.49 & 26.50 & \textbf{0.92}  & 0.0116 \\
    \textbf{MASA-LSTM} & 8.72   & 31.83 & 0.26  & 0.0121 & 13.50  & 19.58 & 0.71  & 0.0087 & 22.12 & 26.61 & 0.90  & 0.0117 \\
    \textbf{MASA-DC}   & 8.70   & 31.77 & 0.25  & 0.0120 & \textbf{15.52}  & \textbf{16.21} & \textbf{0.80}  & \textbf{0.0086} & 14.88 & \textbf{24.29} & 0.60  & \textbf{0.0112} \\
    \hline
    \end{tabular}%
    }
\end{table*}

\noindent\textbf{Performance Analysis}: Table \ref{tab:eval_indexes} reviews the performance of various well-known RL-based approaches against that of the proposed MASA framework using different market observers,  with 
the symbol $\uparrow$ to denote the preference of a larger value in the metrics of AR and SR 
while the symbol $\downarrow$ denoting the favour of a smaller value in MDD and Risk.
From the results of the CSI 300 data set, the AR of the MASA frameworks is at least 1.5\% larger than those of other methods while maintaining the portfolio risks at a relatively low level. 
In particular, the MASA framework integrated with an MLP-based market observer 
achieves the highest AR at 8.87\% and the highest SR at 0.27, 
 thus demonstrating the higher capability of all the proposed agents in the MASA framework to balance the trade-offs among different objectives. 

\begin{figure}[ht]
  \centering
  \includegraphics[width=1\linewidth]{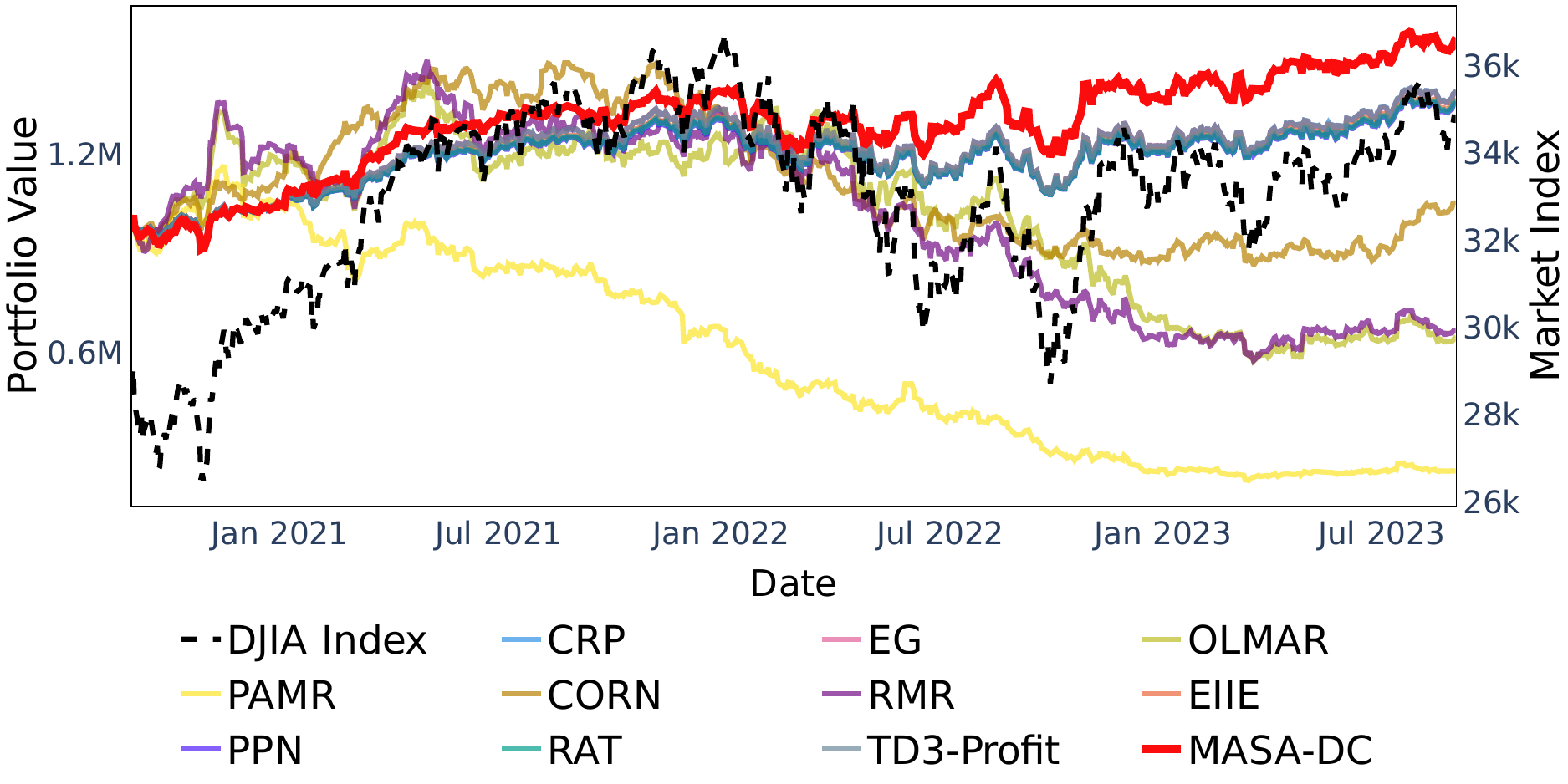}
  \caption{A Comparison of the Portfolio Values of Different Approaches on the DJIA Index}
  \label{fig:pvtrend}
  \Description{Portfolio value trend comparison on the DJIA Index}
\end{figure}

\begin{figure}[ht]
  \centering
  \includegraphics[width=1\linewidth]{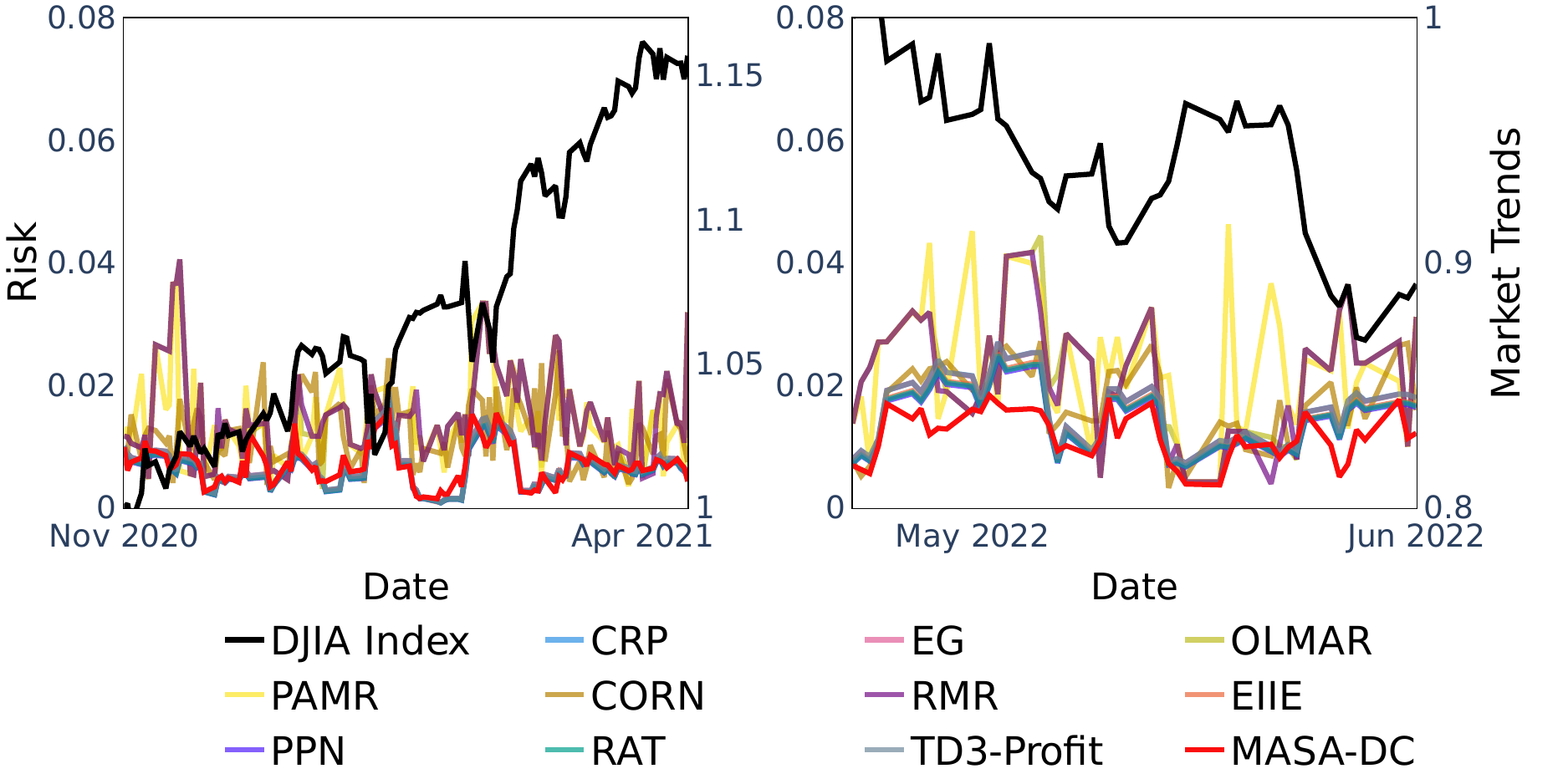}
  \caption{The Risk Comparison on the Uptrend and Downtrend Cases of the DJIA Index}
  \label{fig:riskcase}
  \Description{The Risk Comparison on Uptrend and Downtrend Cases}
\end{figure}

For the attained results on the DJIA index, 
the MASA-DC approach utilising the directional changes (DC) method \cite{tsang2017profiling} as the market observer agent to estimate 
the market trends
significantly outperforms other baseline models in all metrics. 
Specially, the MASA-DC approach attains the AR about 4\% higher than that of the single-agent TD3-Profit approach 
while reducing the maximum possible losses by 3\% when compared to the TD3-Profit in terms of MDD. 
For a clear presentation of the overall results, 
Figure \ref{fig:pvtrend} shows the changes of portfolio values of each approach under evaluation.
In addition, 
Figure \ref{fig:riskcase} shows an interesting example of upward trends in the DJIA market where 
the MASA-DC approach achieves competitive returns while maintaining the potential risks at a relatively lower level 
as compared to those of other approaches.
On the contrary, 
when the financial market stays for long periods of downward trends, 
the portfolio values of those baseline models dramatically decreases as shown in Figure \ref{fig:pvtrend}.
Yet 
the MASA-DC approach can manage to effectively minimise the losses during such adverse market conditions 
when compared to the other approaches.
Figure \ref{fig:riskcase} reveals such a challenging example of downward trends in which 
the short-term risk of the involved portfolio  can be managed well by the MASA-DC approach with less fluctuation even if the market index drops over 10\%, thus  confirming  the effectiveness of the DC-based market observer agent to timely capture the environment changes as the valuable feedback for the solver-based agent to adjust the actions for balancing multiple objectives. 
Furthermore, a similar performance is attained by the MASA approach 
on the other two indexes for which the corresponding graphs of portfolio values and risk comparison can be found in the Appendix
for a more detailed investigation. 

Table \ref{tab:eval_indexes} reviews
the performance of various approaches on the S\&P 500 data set in  which the OLMAR obtains a relatively high AR due to its loser tracking strategy that may typically invest almost the whole capital in a single asset. 
Yet such strategy may not be able to balance the risk in a portfolio and possibly fail in financial markets of downward trends. 
Thus, the OLMAR gets a relatively higher MDD of 68.52\% where it may suffer from huge potential risks. 
Similar to the results obtained on the CSI 300 and DJIA data sets, both the MASA-MLP and MASA-LSTM approaches 
obtain the best performance on balancing the returns and potential risks, achieving a SR of around 0.9 and a MDD of 26\%. 

Moreover, the Wilcoxon rank-sum test \cite{derrac2011practical} is used to compare the statistical significance of the MASA framework against the other approaches with a significance level of 0.05. Clearly, the performance of the MASA is statistically significant against that of other compared approaches on all three data sets except for the specific result attained by OLMAR approach on the S\&P 500 index. Generally speaking, the MASA approach beats all other approaches on the three challenging data sets through the 3 cooperating agents to carefully optimise the possibly conflicting objectives under the highly uncharted environments of 
various financial markets. 

\begin{figure}[ht]
  \centering
  \includegraphics[width=1\linewidth]{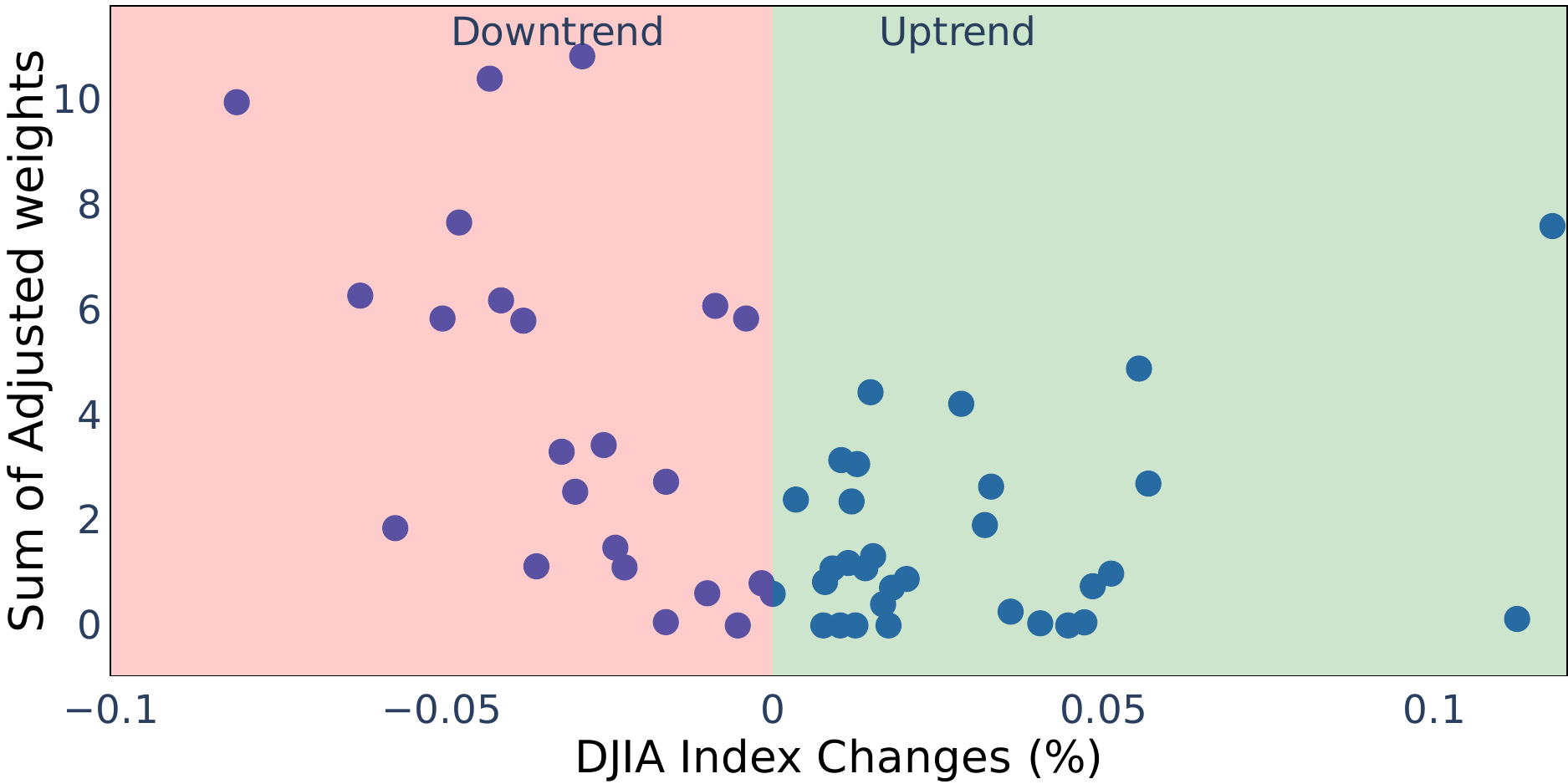}
  \caption{The Contribution of Solver-based Agents on Different Market States}
  \label{fig:distweight}
  \Description{The Contribution of Solver-based Agents on Different Market States}
\end{figure}

\begin{table}[ht]
  \centering
  \caption{The Ablation Study of the Proposed MASA Framework on the CSI 300 Index} 
  \label{tab:ablat_study}
    \resizebox{1\linewidth}{!}{
    \begin{tabular}{c|c|cccc}
    \hline
    \multicolumn{2}{c|}{Models} & \textbf{AR$(\%)\uparrow$}     & \textbf{MDD$(\%)\downarrow$}     & \textbf{SR$\uparrow$}   & \textbf{Risk$\downarrow$} \\
    \hline
    \multirow{3}{*}{Single-Agent} & TD3-Profit    & 7.18 & 33.97 & 0.19 & 0.0128 \\
                                    & TD3-PR         & 7.21 & 33.96 & 0.19 & 0.0128 \\
                                    & TD3-SR         & 7.18 & 33.98 & 0.19 & 0.0128 \\
    \hline
    Dual-Agent                      & MASA-w/oMktObs & 8.17 & 32.71 & 0.23 & 0.0121 \\
    \hline
    \multirow{2}{*}{Triple-Agent} & MASA-MLP & 8.46 & 32.26 & 0.25 & \textbf{0.0119} \\
                                                            & MASA-LSTM      & 8.27 & 32.42 & 0.24 & 0.0121 \\
    (w/o Action Reward)                                       & MASA-DC        & 7.93 & 32.34 & 0.21 & \textbf{0.0119} \\
    \hline
    \multirow{2}{*}{Triple-Agent} & MASA-MLP & \textbf{8.87} & 31.78 & \textbf{0.27} & \textbf{0.0119} \\
                                                             & MASA-LSTM      & 8.72 & 31.83 & 0.26 & 0.0121 \\
    (with Action Reward)                                       & MASA-DC        & 8.70 & \textbf{31.77} & 0.25 & 0.0120 \\
    \hline
    \end{tabular}%
    }
\end{table}
\noindent\textbf{Ablation Study}: 
Table \ref{tab:ablat_study} shows the results of the ablation study of the proposed MASA framework on the CSI 300 index in which
three variants of the TD3-based models are used to compare with the MASA framework utilising the TD3 approach to implement the RL-based agent. 
Specifically, the TD3-Profit model is targeted to maximise total profits while the TD3-PR model combines both profit maximisation and short-term risk minimisation. Besides, the TD3-SR approach uses the SR as the reward function. 
For the dual-agent model, the MASA-w/oMktObs approach combines both the RL-based and solver-based agents to balance the trade-offs of the portfolio optimisation yet there is no market observer agent to provide any additional market information.
For the proposed triple-agent MASA framework, the market observer agent is implemented by the MLP, LSTM and DC method respectively. 
The single-agent approach obtains around 7.20\% of returns per year yet with potential losses of 33\% during the trading period. 
Moreover, the MASA-w/oMktObs efficiently reduces the investment risks to avoid great losses even when no extra information about market changes is provided. Thus, the total AR of the dual-agent model is increased by 1\% against that of single-agent model. Meanwhile,  a relatively higher SR is achieved by the MASA-w/oMktObs due to a better trade-off between returns and risks.
 Furthermore, the MASA can better estimate the potential risks while pursuing higher returns after 
 considering more latest market information from the market observer agent. 
 The resulting risks and MDD are dropped to 0.0119 and 31.7\% respectively, 
 with some further improvement on both the total returns and SR. 
 To demonstrate the effectiveness of the solver-based agent on the risk management, 
 Figure~\ref{fig:distweight} shows the relationship between market state changes and the sum of weights adjusted by the solver-based agent. Clearly, the solver-based agent makes larger weighting adjustments to manage risks at a relatively low level 
 after considering the market information provided by the market observer, 
 especially when the potential risks are increased sharply over successive episodes. 
Obviously, the ablation studies confirm the contributions of both the solver-based and market observer agents in the proposed MASA framework that can effectively tackle the trade-offs between different objectives under the 
highly volatile environments
of financial markets.

Table \ref{tab:ablat_study} shows the effectiveness of the reward of the action generated by the RL-based agent in the proposed MASA framework. The MASA variant without considering the reward of the action by the RL-based agent
 still performs better than the single-agent or dual-agent framework in balancing the profits and risks 
 especially when the MLP or LSTM model is used for the market observer agent. 
 When considering the rewards of generated actions, the risk-aware information provided by the solver-based agent can guide the policy of the RL-based agent toward higher profits and less potential risks for which 
 the MASA framework can enhance the AR by 0.5\% and the MDD by 1\%.

Furthermore,
the top 20 and 30 stocks of each index are selected to study the scalability of the MASA framework on large-scale portfolios,  
except for the CSI 300 index due to the limited data sources. When constructing a portfolio of 20 assets in the DJIA market, the MASA-MLP achieves the highest SR of 0.80 and the highest AR of 14\% while the best baseline approach obtains a SR of 0.61 and a AR of 11\% only.  After increasing the portfolio size to 30 assets, the MASA framework still has a significant improvement against those of other approaches on all metrics. Similar results are obtained by the MASA framework in the S\&P 500 market, that can be found in the Appendix. Undoubtedly, all the obtained results validate that the MASA framework can achieve a better performance in balancing different goals when the problem size increases.

\section{Concluding Remarks} \label{sec:concl} 
In recent years, deep or reinforcement learning (RL) approaches have been adapted as intelligent and reactive agents to quickly learn and respond with newly revised investment strategies for portfolio management under the highly volatile financial market environments particularly under the threats of global or regional conflicts, pandemics, natural disasters, etc.   
Yet due to the very complex correlations among different financial sectors, and the fluctuating trends in various financial markets, a deep or reinforcement learning based agent can be biased in maximising the total returns of the newly formulated investment portfolio while neglecting its potential risks under the turmoil of various market conditions in the global or regional sectors. 
Accordingly, a multi-agent and self-adaptive framework namely the MASA is proposed in which a sophisticated multi-agent reinforcement learning approach is adopted through both of the RL-based and solver-based agents working
to carefully and dynamically balance the trade-off between the overall portfolio returns and their potential risks. 
In addition, a very flexible and proactive agent as the market observer is integrated into the proposed MASA framework to provide  the estimated market conditions and trends as additional information for multi-agent RL approach to carefully consider so as to quickly adapt to the ever-changing market conditions. 

To demonstrated the potential advantages of our proposal, 
a prototype of the proposed MASA framework is evaluated 
against various well-known RL-based approaches on the challenging data sets of the CSI 300, Dow Jones Industrial Average and S\&P 500 indexes over the past 10 years.
The obtained empirical results clearly reveal the remarkable performance of our proposed MASA framework based on the multi-agent RL approach when compared against those of other well-known RL-based approaches on the 3 data sets of widely recognised financial indexes in China and the United States. 
More importantly, our proposed MASA framework shed lights on many possible directions for future investigation.
First, the thorough investigation on using different meta-heuristic based optimisers 
such as the evolutionary algorithms or the PSO for the solver-based agent should be interesting.
Besides, experimenting various intelligent approaches for the market observer agent is worth exploring.
Last but not least, 
the potential applications of the proposed MASA model for various resource allocation, planning or disaster recovery 
in which the risk management is critical and timely 
should be very valuable for our future studies.



\begin{acks}
The authors wish to express our deepest gratitude to Professor Edward Tsang for his fruitful discussion on this work, and also the anonymous reviewers for their valuable feedback.
\end{acks}

\balance


\bibliographystyle{ACM-Reference-Format} 
\bibliography{MASA}


\begin{thebibliography}{47}


\ifx \showCODEN    \undefined \def \showCODEN     #1{\unskip}     \fi
\ifx \showDOI      \undefined \def \showDOI       #1{#1}\fi
\ifx \showISBNx    \undefined \def \showISBNx     #1{\unskip}     \fi
\ifx \showISBNxiii \undefined \def \showISBNxiii  #1{\unskip}     \fi
\ifx \showISSN     \undefined \def \showISSN      #1{\unskip}     \fi
\ifx \showLCCN     \undefined \def \showLCCN      #1{\unskip}     \fi
\ifx \shownote     \undefined \def \shownote      #1{#1}          \fi
\ifx \showarticletitle \undefined \def \showarticletitle #1{#1}   \fi
\ifx \showURL      \undefined \def \showURL       {\relax}        \fi
\providecommand\bibfield[2]{#2}
\providecommand\bibinfo[2]{#2}
\providecommand\natexlab[1]{#1}
\providecommand\showeprint[2][]{arXiv:#2}

\bibitem[\protect\citeauthoryear{Alfaro and Sagner}{Alfaro and Sagner}{2010}]%
        {alfaro2010financial}
\bibfield{author}{\bibinfo{person}{Rodrigo Alfaro} {and} \bibinfo{person}{Andres Sagner}.} \bibinfo{year}{2010}\natexlab{}.
\newblock \showarticletitle{Financial Forecast for the Relative Strength Index}.
\newblock  (\bibinfo{year}{2010}).
\newblock


\bibitem[\protect\citeauthoryear{Bakhach, Tsang, and Jalalian}{Bakhach et~al\mbox{.}}{2016}]%
        {bakhach2016forecasting}
\bibfield{author}{\bibinfo{person}{Amer Bakhach}, \bibinfo{person}{Edward~PK Tsang}, {and} \bibinfo{person}{Hamid Jalalian}.} \bibinfo{year}{2016}\natexlab{}.
\newblock \showarticletitle{Forecasting directional changes in the fx markets}. In \bibinfo{booktitle}{\emph{2016 IEEE Symposium Series on Computational Intelligence (SSCI)}}. IEEE, \bibinfo{pages}{1--8}.
\newblock


\bibitem[\protect\citeauthoryear{Belcak, Calliess, and Zohren}{Belcak et~al\mbox{.}}{2022}]%
        {belcak_2022}
\bibfield{author}{\bibinfo{person}{Peter Belcak}, \bibinfo{person}{Jan-Peter Calliess}, {and} \bibinfo{person}{Stefan Zohren}.} \bibinfo{year}{2022}\natexlab{}.
\newblock \showarticletitle{Fast Agent-Based Simulation Framework with Applications to Reinforcement Learning and the Study of Trading Latency Effects}. In \bibinfo{booktitle}{\emph{Multi-Agent-Based Simulation XXII}}, \bibfield{editor}{\bibinfo{person}{Koen~H. Van~Dam} {and} \bibinfo{person}{Nicolas Verstaevel}} (Eds.). \bibinfo{publisher}{Springer International Publishing}, \bibinfo{address}{Cham}, \bibinfo{pages}{42--56}.
\newblock


\bibitem[\protect\citeauthoryear{Calliess and Zohren}{Calliess and Zohren}{2021}]%
        {Calliess_2021}
\bibfield{author}{\bibinfo{person}{Jan-Peter Calliess} {and} \bibinfo{person}{Stefan Zohren}.} \bibinfo{year}{2021}\natexlab{}.
\newblock \showarticletitle{Agent-Based Models in Finance and Market Simulations}.
\newblock  (\bibinfo{year}{2021}).
\newblock
\urldef\tempurl%
\url{https://oxford-man.ox.ac.uk/projects/agent-based-models-in-finance-and-market-simulations/}
\showURL{%
\tempurl}


\bibitem[\protect\citeauthoryear{Cover}{Cover}{1991}]%
        {cover1991universal}
\bibfield{author}{\bibinfo{person}{Thomas~M Cover}.} \bibinfo{year}{1991}\natexlab{}.
\newblock \showarticletitle{Universal Portfolios}.
\newblock \bibinfo{journal}{\emph{Mathematical finance}} (\bibinfo{year}{1991}).
\newblock


\bibitem[\protect\citeauthoryear{Cuschieri, Vella, and Bajada}{Cuschieri et~al\mbox{.}}{2021}]%
        {cuschieri2021td3}
\bibfield{author}{\bibinfo{person}{Nigel Cuschieri}, \bibinfo{person}{Vince Vella}, {and} \bibinfo{person}{Josef Bajada}.} \bibinfo{year}{2021}\natexlab{}.
\newblock \showarticletitle{TD3-Based Ensemble Reinforcement Learning for Financial Portfolio Optimisation}.
\newblock \bibinfo{journal}{\emph{FinPlan 2021}} (\bibinfo{year}{2021}), \bibinfo{pages}{6}.
\newblock


\bibitem[\protect\citeauthoryear{Derrac, Garc{\'\i}a, Molina, and Herrera}{Derrac et~al\mbox{.}}{2011}]%
        {derrac2011practical}
\bibfield{author}{\bibinfo{person}{Joaqu{\'\i}n Derrac}, \bibinfo{person}{Salvador Garc{\'\i}a}, \bibinfo{person}{Daniel Molina}, {and} \bibinfo{person}{Francisco Herrera}.} \bibinfo{year}{2011}\natexlab{}.
\newblock \showarticletitle{A practical tutorial on the use of nonparametric statistical tests as a methodology for comparing evolutionary and swarm intelligence algorithms}.
\newblock \bibinfo{journal}{\emph{Swarm and Evolutionary Computation}} \bibinfo{volume}{1}, \bibinfo{number}{1} (\bibinfo{year}{2011}), \bibinfo{pages}{3--18}.
\newblock


\bibitem[\protect\citeauthoryear{Dre{\.z}ewski and Doroz}{Dre{\.z}ewski and Doroz}{2017}]%
        {drezewski2017agent}
\bibfield{author}{\bibinfo{person}{Rafa{\l} Dre{\.z}ewski} {and} \bibinfo{person}{Krzysztof Doroz}.} \bibinfo{year}{2017}\natexlab{}.
\newblock \showarticletitle{An agent-based co-evolutionary multi-objective algorithm for portfolio optimization}.
\newblock \bibinfo{journal}{\emph{Symmetry}} \bibinfo{volume}{9}, \bibinfo{number}{9} (\bibinfo{year}{2017}), \bibinfo{pages}{168}.
\newblock


\bibitem[\protect\citeauthoryear{Duan, Wang, Zhang, and Li}{Duan et~al\mbox{.}}{2022}]%
        {duan2022factorvae}
\bibfield{author}{\bibinfo{person}{Yitong Duan}, \bibinfo{person}{Lei Wang}, \bibinfo{person}{Qizhong Zhang}, {and} \bibinfo{person}{Jian Li}.} \bibinfo{year}{2022}\natexlab{}.
\newblock \showarticletitle{FactorVAE: A Probabilistic Dynamic Factor Model Based on Variational Autoencoder for Predicting Cross-sectional Stock Returns}. In \bibinfo{booktitle}{\emph{Proceedings of the AAAI Conference on Artificial Intelligence}}.
\newblock


\bibitem[\protect\citeauthoryear{Fujimoto, Hoof, and Meger}{Fujimoto et~al\mbox{.}}{2018}]%
        {fujimoto2018addressing}
\bibfield{author}{\bibinfo{person}{Scott Fujimoto}, \bibinfo{person}{Herke Hoof}, {and} \bibinfo{person}{David Meger}.} \bibinfo{year}{2018}\natexlab{}.
\newblock \showarticletitle{Addressing Function Approximation Error in Actor-critic Methods}. In \bibinfo{booktitle}{\emph{Proceedings of the International Conference on Machine Learning}}. PMLR.
\newblock


\bibitem[\protect\citeauthoryear{Goodell, Kumar, Lim, and Pattnaik}{Goodell et~al\mbox{.}}{2021}]%
        {goodell2021artificial}
\bibfield{author}{\bibinfo{person}{John~W Goodell}, \bibinfo{person}{Satish Kumar}, \bibinfo{person}{Weng~Marc Lim}, {and} \bibinfo{person}{Debidutta Pattnaik}.} \bibinfo{year}{2021}\natexlab{}.
\newblock \showarticletitle{Artificial intelligence and machine learning in finance: Identifying foundations, themes, and research clusters from bibliometric analysis}.
\newblock \bibinfo{journal}{\emph{Journal of Behavioral and Experimental Finance}}  \bibinfo{volume}{32} (\bibinfo{year}{2021}), \bibinfo{pages}{100577}.
\newblock


\bibitem[\protect\citeauthoryear{Gunjan and Bhattacharyya}{Gunjan and Bhattacharyya}{2023}]%
        {gunjan2023brief}
\bibfield{author}{\bibinfo{person}{Abhishek Gunjan} {and} \bibinfo{person}{Siddhartha Bhattacharyya}.} \bibinfo{year}{2023}\natexlab{}.
\newblock \showarticletitle{A brief review of portfolio optimization techniques}.
\newblock \bibinfo{journal}{\emph{Artificial Intelligence Review}} \bibinfo{volume}{56}, \bibinfo{number}{5} (\bibinfo{year}{2023}), \bibinfo{pages}{3847--3886}.
\newblock


\bibitem[\protect\citeauthoryear{Hafezi, Shahrabi, and Hadavandi}{Hafezi et~al\mbox{.}}{2015}]%
        {hafezi2015bat}
\bibfield{author}{\bibinfo{person}{Reza Hafezi}, \bibinfo{person}{Jamal Shahrabi}, {and} \bibinfo{person}{Esmaeil Hadavandi}.} \bibinfo{year}{2015}\natexlab{}.
\newblock \showarticletitle{A bat-neural network multi-agent system (BNNMAS) for stock price prediction: Case study of DAX stock price}.
\newblock \bibinfo{journal}{\emph{Applied Soft Computing}}  \bibinfo{volume}{29} (\bibinfo{year}{2015}), \bibinfo{pages}{196--210}.
\newblock


\bibitem[\protect\citeauthoryear{Helmbold, Schapire, Singer, and Warmuth}{Helmbold et~al\mbox{.}}{1998}]%
        {helmbold1998line}
\bibfield{author}{\bibinfo{person}{David~P Helmbold}, \bibinfo{person}{Robert~E Schapire}, \bibinfo{person}{Yoram Singer}, {and} \bibinfo{person}{Manfred~K Warmuth}.} \bibinfo{year}{1998}\natexlab{}.
\newblock \showarticletitle{On-line Portfolio Selection Using Multiplicative Updates}.
\newblock \bibinfo{journal}{\emph{Mathematical Finance}} (\bibinfo{year}{1998}).
\newblock


\bibitem[\protect\citeauthoryear{Hochreiter and Schmidhuber}{Hochreiter and Schmidhuber}{1997}]%
        {hochreiter1997long}
\bibfield{author}{\bibinfo{person}{Sepp Hochreiter} {and} \bibinfo{person}{J{\"u}rgen Schmidhuber}.} \bibinfo{year}{1997}\natexlab{}.
\newblock \showarticletitle{Long short-term memory}.
\newblock \bibinfo{journal}{\emph{Neural computation}} \bibinfo{volume}{9}, \bibinfo{number}{8} (\bibinfo{year}{1997}), \bibinfo{pages}{1735--1780}.
\newblock


\bibitem[\protect\citeauthoryear{Huang, Zhou, Li, Hoi, and Zhou}{Huang et~al\mbox{.}}{2016}]%
        {huang2016robust}
\bibfield{author}{\bibinfo{person}{Dingjiang Huang}, \bibinfo{person}{Junlong Zhou}, \bibinfo{person}{Bin Li}, \bibinfo{person}{Steven~CH Hoi}, {and} \bibinfo{person}{Shuigeng Zhou}.} \bibinfo{year}{2016}\natexlab{}.
\newblock \showarticletitle{Robust median reversion strategy for online portfolio selection}.
\newblock \bibinfo{journal}{\emph{IEEE Transactions on Knowledge and Data Engineering}} \bibinfo{volume}{28}, \bibinfo{number}{9} (\bibinfo{year}{2016}), \bibinfo{pages}{2480--2493}.
\newblock


\bibitem[\protect\citeauthoryear{Jiang, Xu, and Liang}{Jiang et~al\mbox{.}}{2017}]%
        {jiang2017deep}
\bibfield{author}{\bibinfo{person}{Zhengyao Jiang}, \bibinfo{person}{Dixing Xu}, {and} \bibinfo{person}{Jinjun Liang}.} \bibinfo{year}{2017}\natexlab{}.
\newblock \showarticletitle{A Deep Reinforcement Learning Framework for the Financial Portfolio Management Problem}.
\newblock \bibinfo{journal}{\emph{arXiv preprint arXiv:1706.10059}} (\bibinfo{year}{2017}).
\newblock


\bibitem[\protect\citeauthoryear{Kampouridis, Kanellopoulos, Kyropoulou, Melissourgos, and Voudouris}{Kampouridis et~al\mbox{.}}{2022}]%
        {Kampouridis_2022}
\bibfield{author}{\bibinfo{person}{Michael Kampouridis}, \bibinfo{person}{Panagiotis Kanellopoulos}, \bibinfo{person}{Maria Kyropoulou}, \bibinfo{person}{Themistoklis Melissourgos}, {and} \bibinfo{person}{Alexandros~A. Voudouris}.} \bibinfo{year}{2022}\natexlab{}.
\newblock \showarticletitle{Multi-agent systems for computational economics and finance}.
\newblock \bibinfo{journal}{\emph{{AI} Communications}} \bibinfo{volume}{35}, \bibinfo{number}{4} (\bibinfo{date}{sep} \bibinfo{year}{2022}), \bibinfo{pages}{369--380}.
\newblock
\urldef\tempurl%
\url{https://doi.org/10.3233/aic-220117}
\showDOI{\tempurl}


\bibitem[\protect\citeauthoryear{Kennedy and Eberhart}{Kennedy and Eberhart}{1995}]%
        {kennedy1995particle}
\bibfield{author}{\bibinfo{person}{James Kennedy} {and} \bibinfo{person}{Russell Eberhart}.} \bibinfo{year}{1995}\natexlab{}.
\newblock \showarticletitle{Particle swarm optimization}. In \bibinfo{booktitle}{\emph{Proceedings of ICNN'95-international conference on neural networks}}, Vol.~\bibinfo{volume}{4}. IEEE, \bibinfo{pages}{1942--1948}.
\newblock


\bibitem[\protect\citeauthoryear{Kolm and Ritter}{Kolm and Ritter}{2020}]%
        {kolm2020modern}
\bibfield{author}{\bibinfo{person}{Petter~N Kolm} {and} \bibinfo{person}{Gordon Ritter}.} \bibinfo{year}{2020}\natexlab{}.
\newblock \showarticletitle{Modern perspectives on reinforcement learning in finance}.
\newblock \bibinfo{journal}{\emph{Modern Perspectives on Reinforcement Learning in Finance (September 6, 2019). The Journal of Machine Learning in Finance}} \bibinfo{volume}{1}, \bibinfo{number}{1} (\bibinfo{year}{2020}).
\newblock


\bibitem[\protect\citeauthoryear{Li and Hoi}{Li and Hoi}{2012}]%
        {li26line}
\bibfield{author}{\bibinfo{person}{Bin Li} {and} \bibinfo{person}{Steven~CH Hoi}.} \bibinfo{year}{2012}\natexlab{}.
\newblock \showarticletitle{On-Line Portfolio Selection with Moving Average Reversion}. In \bibinfo{booktitle}{\emph{Proceedings of the International Conference on Machine Learning}}. PMLR.
\newblock


\bibitem[\protect\citeauthoryear{Li and Hoi}{Li and Hoi}{2014}]%
        {li2014online}
\bibfield{author}{\bibinfo{person}{Bin Li} {and} \bibinfo{person}{Steven~CH Hoi}.} \bibinfo{year}{2014}\natexlab{}.
\newblock \showarticletitle{Online Portfolio Selection: A Survey}.
\newblock \bibinfo{journal}{\emph{ACM Computing Surveys (CSUR)}} (\bibinfo{year}{2014}).
\newblock


\bibitem[\protect\citeauthoryear{Li, Hoi, and Gopalkrishnan}{Li et~al\mbox{.}}{2011}]%
        {li2011corn}
\bibfield{author}{\bibinfo{person}{Bin Li}, \bibinfo{person}{Steven~CH Hoi}, {and} \bibinfo{person}{Vivekanand Gopalkrishnan}.} \bibinfo{year}{2011}\natexlab{}.
\newblock \showarticletitle{Corn: Correlation-driven Nonparametric Learning Approach for Portfolio Selection}.
\newblock \bibinfo{journal}{\emph{ACM Transactions on Intelligent Systems and Technology}} (\bibinfo{year}{2011}).
\newblock


\bibitem[\protect\citeauthoryear{Li, Zhao, Hoi, and Gopalkrishnan}{Li et~al\mbox{.}}{2012}]%
        {li2012pamr}
\bibfield{author}{\bibinfo{person}{Bin Li}, \bibinfo{person}{Peilin Zhao}, \bibinfo{person}{Steven~CH Hoi}, {and} \bibinfo{person}{Vivekanand Gopalkrishnan}.} \bibinfo{year}{2012}\natexlab{}.
\newblock \showarticletitle{PAMR: Passive Aggressive Mean Reversion Strategy for Portfolio Selection}.
\newblock \bibinfo{journal}{\emph{Machine learning}} (\bibinfo{year}{2012}).
\newblock


\bibitem[\protect\citeauthoryear{Liang, Zhu, Zheng, and Wang}{Liang et~al\mbox{.}}{2021}]%
        {liang2021adaptive}
\bibfield{author}{\bibinfo{person}{Qianqiao Liang}, \bibinfo{person}{Mengying Zhu}, \bibinfo{person}{Xiaolin Zheng}, {and} \bibinfo{person}{Yan Wang}.} \bibinfo{year}{2021}\natexlab{}.
\newblock \showarticletitle{An Adaptive News-Driven Method for CVaR-sensitive Online Portfolio Selection in Non-Stationary Financial Markets.}. In \bibinfo{booktitle}{\emph{Proceedings of the IJCAI Conference on Artificial Intelligence}}.
\newblock


\bibitem[\protect\citeauthoryear{Lin}{Lin}{1991}]%
        {lin1991}
\bibfield{author}{\bibinfo{person}{J. Lin}.} \bibinfo{year}{1991}\natexlab{}.
\newblock \showarticletitle{Divergence measures based on the Shannon entropy}.
\newblock \bibinfo{journal}{\emph{IEEE Transactions on Information Theory}} \bibinfo{volume}{37}, \bibinfo{number}{1} (\bibinfo{year}{1991}), \bibinfo{pages}{145--151}.
\newblock
\urldef\tempurl%
\url{https://doi.org/10.1109/18.61115}
\showDOI{\tempurl}


\bibitem[\protect\citeauthoryear{Liu, Liu, Zhao, Pan, and Liu}{Liu et~al\mbox{.}}{2020}]%
        {liu2020adaptive}
\bibfield{author}{\bibinfo{person}{Yang Liu}, \bibinfo{person}{Qi Liu}, \bibinfo{person}{Hongke Zhao}, \bibinfo{person}{Zhen Pan}, {and} \bibinfo{person}{Chuanren Liu}.} \bibinfo{year}{2020}\natexlab{}.
\newblock \showarticletitle{Adaptive Quantitative Trading: An Imitative Deep Reinforcement Learning Approach}. In \bibinfo{booktitle}{\emph{Proceedings of the AAAI conference on artificial intelligence}}.
\newblock


\bibitem[\protect\citeauthoryear{Maglaras, Moallemi, and Wang}{Maglaras et~al\mbox{.}}{2022}]%
        {maglaras2022deep}
\bibfield{author}{\bibinfo{person}{Costis Maglaras}, \bibinfo{person}{Ciamac~C Moallemi}, {and} \bibinfo{person}{Muye Wang}.} \bibinfo{year}{2022}\natexlab{}.
\newblock \showarticletitle{A deep learning approach to estimating fill probabilities in a limit order book}.
\newblock \bibinfo{journal}{\emph{Quantitative Finance}} \bibinfo{volume}{22}, \bibinfo{number}{11} (\bibinfo{year}{2022}), \bibinfo{pages}{1989--2003}.
\newblock


\bibitem[\protect\citeauthoryear{Markowitz and Todd}{Markowitz and Todd}{2000}]%
        {markowitz2000mean}
\bibfield{author}{\bibinfo{person}{Harry~M Markowitz} {and} \bibinfo{person}{G~Peter Todd}.} \bibinfo{year}{2000}\natexlab{}.
\newblock \bibinfo{booktitle}{\emph{Mean-variance Analysis in Portfolio Choice and Capital Markets}}.
\newblock \bibinfo{publisher}{John Wiley \& Sons}.
\newblock


\bibitem[\protect\citeauthoryear{Menéndez, Pardo, Pardo, and Pardo}{Menéndez et~al\mbox{.}}{1997}]%
        {MENENDEZ1997307}
\bibfield{author}{\bibinfo{person}{M.L. Menéndez}, \bibinfo{person}{J.A. Pardo}, \bibinfo{person}{L. Pardo}, {and} \bibinfo{person}{M.C. Pardo}.} \bibinfo{year}{1997}\natexlab{}.
\newblock \showarticletitle{The Jensen-Shannon divergence}.
\newblock \bibinfo{journal}{\emph{Journal of the Franklin Institute}} \bibinfo{volume}{334}, \bibinfo{number}{2} (\bibinfo{year}{1997}), \bibinfo{pages}{307--318}.
\newblock
\showISSN{0016-0032}
\urldef\tempurl%
\url{https://doi.org/10.1016/S0016-0032(96)00063-4}
\showDOI{\tempurl}


\bibitem[\protect\citeauthoryear{Nabipour, Nayyeri, Jabani, Shahab, and Mosavi}{Nabipour et~al\mbox{.}}{2020}]%
        {nabipour2020predicting}
\bibfield{author}{\bibinfo{person}{Mojtaba Nabipour}, \bibinfo{person}{Pooyan Nayyeri}, \bibinfo{person}{Hamed Jabani}, \bibinfo{person}{S Shahab}, {and} \bibinfo{person}{Amir Mosavi}.} \bibinfo{year}{2020}\natexlab{}.
\newblock \showarticletitle{Predicting stock market trends using machine learning and deep learning algorithms via continuous and binary data; a comparative analysis}.
\newblock \bibinfo{journal}{\emph{IEEE Access}}  \bibinfo{volume}{8} (\bibinfo{year}{2020}), \bibinfo{pages}{150199--150212}.
\newblock


\bibitem[\protect\citeauthoryear{Puterman}{Puterman}{1990}]%
        {puterman1990markov}
\bibfield{author}{\bibinfo{person}{Martin~L Puterman}.} \bibinfo{year}{1990}\natexlab{}.
\newblock \showarticletitle{Markov decision processes}.
\newblock \bibinfo{journal}{\emph{Handbooks in operations research and management science}}  \bibinfo{volume}{2} (\bibinfo{year}{1990}), \bibinfo{pages}{331--434}.
\newblock


\bibitem[\protect\citeauthoryear{Raudys, Len{\v{c}}iauskas, and Mal{\v{c}}ius}{Raudys et~al\mbox{.}}{2013}]%
        {raudys2013moving}
\bibfield{author}{\bibinfo{person}{Aistis Raudys}, \bibinfo{person}{Vaidotas Len{\v{c}}iauskas}, {and} \bibinfo{person}{Edmundas Mal{\v{c}}ius}.} \bibinfo{year}{2013}\natexlab{}.
\newblock \showarticletitle{Moving averages for financial data smoothing}. In \bibinfo{booktitle}{\emph{Information and Software Technologies: 19th International Conference, ICIST 2013, Kaunas, Lithuania, October 2013. Proceedings 19}}. Springer, \bibinfo{pages}{34--45}.
\newblock


\bibitem[\protect\citeauthoryear{Sawik}{Sawik}{2012}]%
        {sawik2012bi}
\bibfield{author}{\bibinfo{person}{Bartosz Sawik}.} \bibinfo{year}{2012}\natexlab{}.
\newblock \showarticletitle{Bi-criteria portfolio optimization models with percentile and symmetric risk measures by mathematical programming}.
\newblock \bibinfo{journal}{\emph{Przeglad Elektrotechniczny}} \bibinfo{volume}{88}, \bibinfo{number}{10B} (\bibinfo{year}{2012}), \bibinfo{pages}{176--180}.
\newblock


\bibitem[\protect\citeauthoryear{Sivaram, Lydia, Pustokhina, Pustokhin, Elhoseny, Joshi, and Shankar}{Sivaram et~al\mbox{.}}{2020}]%
        {sivaram2020optimal}
\bibfield{author}{\bibinfo{person}{M Sivaram}, \bibinfo{person}{E~Laxmi Lydia}, \bibinfo{person}{Irina~V Pustokhina}, \bibinfo{person}{Denis~Alexandrovich Pustokhin}, \bibinfo{person}{Mohamed Elhoseny}, \bibinfo{person}{Gyanendra~Prasad Joshi}, {and} \bibinfo{person}{K Shankar}.} \bibinfo{year}{2020}\natexlab{}.
\newblock \showarticletitle{An optimal least square support vector machine based earnings prediction of blockchain financial products}.
\newblock \bibinfo{journal}{\emph{IEEE Access}}  \bibinfo{volume}{8} (\bibinfo{year}{2020}), \bibinfo{pages}{120321--120330}.
\newblock


\bibitem[\protect\citeauthoryear{Storn and Price}{Storn and Price}{1997}]%
        {storn1997differential}
\bibfield{author}{\bibinfo{person}{Rainer Storn} {and} \bibinfo{person}{Kenneth Price}.} \bibinfo{year}{1997}\natexlab{}.
\newblock \showarticletitle{Differential evolution--a simple and efficient heuristic for global optimization over continuous spaces}.
\newblock \bibinfo{journal}{\emph{Journal of global optimization}} \bibinfo{volume}{11}, \bibinfo{number}{4} (\bibinfo{year}{1997}), \bibinfo{pages}{341--359}.
\newblock


\bibitem[\protect\citeauthoryear{Tsang}{Tsang}{2023}]%
        {tsang2023ai}
\bibfield{author}{\bibinfo{person}{Edward~PK Tsang}.} \bibinfo{year}{2023}\natexlab{}.
\newblock \bibinfo{booktitle}{\emph{AI for Finance}}.
\newblock \bibinfo{publisher}{CRC Press}.
\newblock


\bibitem[\protect\citeauthoryear{Tsang, Tao, Serguieva, and Ma}{Tsang et~al\mbox{.}}{2017}]%
        {tsang2017profiling}
\bibfield{author}{\bibinfo{person}{Edward~PK Tsang}, \bibinfo{person}{Ran Tao}, \bibinfo{person}{Antoaneta Serguieva}, {and} \bibinfo{person}{Shuai Ma}.} \bibinfo{year}{2017}\natexlab{}.
\newblock \showarticletitle{Profiling high-frequency equity price movements in directional changes}.
\newblock \bibinfo{journal}{\emph{Quantitative finance}} \bibinfo{volume}{17}, \bibinfo{number}{2} (\bibinfo{year}{2017}), \bibinfo{pages}{217--225}.
\newblock


\bibitem[\protect\citeauthoryear{Wang, Li, Wang, and Zheng}{Wang et~al\mbox{.}}{2021b}]%
        {wang2021hierarchical}
\bibfield{author}{\bibinfo{person}{Heyuan Wang}, \bibinfo{person}{Shun Li}, \bibinfo{person}{Tengjiao Wang}, {and} \bibinfo{person}{Jiayi Zheng}.} \bibinfo{year}{2021}\natexlab{b}.
\newblock \showarticletitle{Hierarchical Adaptive Temporal-Relational Modeling for Stock Trend Prediction.}. In \bibinfo{booktitle}{\emph{Proceedings of the IJCAI Conference on Artificial Intelligence}}.
\newblock


\bibitem[\protect\citeauthoryear{Wang, Shivanna, Cheng, Jain, Lin, Hong, and Chi}{Wang et~al\mbox{.}}{2021c}]%
        {wang2021dcn}
\bibfield{author}{\bibinfo{person}{Ruoxi Wang}, \bibinfo{person}{Rakesh Shivanna}, \bibinfo{person}{Derek Cheng}, \bibinfo{person}{Sagar Jain}, \bibinfo{person}{Dong Lin}, \bibinfo{person}{Lichan Hong}, {and} \bibinfo{person}{Ed Chi}.} \bibinfo{year}{2021}\natexlab{c}.
\newblock \showarticletitle{Dcn v2: Improved deep \& cross network and practical lessons for web-scale learning to rank systems}. In \bibinfo{booktitle}{\emph{Proceedings of the web conference 2021}}. \bibinfo{pages}{1785--1797}.
\newblock


\bibitem[\protect\citeauthoryear{Wang, Huang, Tu, Zhang, and Xu}{Wang et~al\mbox{.}}{2021a}]%
        {wang2021deeptrader}
\bibfield{author}{\bibinfo{person}{Zhicheng Wang}, \bibinfo{person}{Biwei Huang}, \bibinfo{person}{Shikui Tu}, \bibinfo{person}{Kun Zhang}, {and} \bibinfo{person}{Lei Xu}.} \bibinfo{year}{2021}\natexlab{a}.
\newblock \showarticletitle{DeepTrader: a deep reinforcement learning approach for risk-return balanced portfolio management with market conditions Embedding}. In \bibinfo{booktitle}{\emph{Proceedings of the AAAI conference on artificial intelligence}}, Vol.~\bibinfo{volume}{35}. \bibinfo{pages}{643--650}.
\newblock


\bibitem[\protect\citeauthoryear{Xu, Zhang, Ye, Zhao, and Tan}{Xu et~al\mbox{.}}{2021}]%
        {xu2021relation}
\bibfield{author}{\bibinfo{person}{Ke Xu}, \bibinfo{person}{Yifan Zhang}, \bibinfo{person}{Deheng Ye}, \bibinfo{person}{Peilin Zhao}, {and} \bibinfo{person}{Mingkui Tan}.} \bibinfo{year}{2021}\natexlab{}.
\newblock \showarticletitle{Relation-aware Transformer for Portfolio Policy Learning}. In \bibinfo{booktitle}{\emph{Proceedings of the IJCAI Conference on Artificial Intelligence}}.
\newblock


\bibitem[\protect\citeauthoryear{Yang, Zheng, Liang, Han, and Zhu}{Yang et~al\mbox{.}}{2022}]%
        {yang2022smart}
\bibfield{author}{\bibinfo{person}{Mengyuan Yang}, \bibinfo{person}{Xiaolin Zheng}, \bibinfo{person}{Qianqiao Liang}, \bibinfo{person}{Bing Han}, {and} \bibinfo{person}{Mengying Zhu}.} \bibinfo{year}{2022}\natexlab{}.
\newblock \showarticletitle{A Smart Trader for Portfolio Management based on Normalizing Flows}. In \bibinfo{booktitle}{\emph{Proceedings of the IJCAI Conference on Artificial Intelligence}}.
\newblock


\bibitem[\protect\citeauthoryear{Ye, Pei, Wang, Chen, Zhu, Xiao, and Li}{Ye et~al\mbox{.}}{2020}]%
        {ye2020reinforcement}
\bibfield{author}{\bibinfo{person}{Yunan Ye}, \bibinfo{person}{Hengzhi Pei}, \bibinfo{person}{Boxin Wang}, \bibinfo{person}{Pin-Yu Chen}, \bibinfo{person}{Yada Zhu}, \bibinfo{person}{Ju Xiao}, {and} \bibinfo{person}{Bo Li}.} \bibinfo{year}{2020}\natexlab{}.
\newblock \showarticletitle{Reinforcement-learning Based Portfolio Management with Augmented Asset Movement Prediction States}. In \bibinfo{booktitle}{\emph{Proceedings of the AAAI Conference on Artificial Intelligence}}.
\newblock


\bibitem[\protect\citeauthoryear{Zhang, Li, and Li}{Zhang et~al\mbox{.}}{2020a}]%
        {ZHANG2020206}
\bibfield{author}{\bibinfo{person}{Fengjiao Zhang}, \bibinfo{person}{Jie Li}, {and} \bibinfo{person}{Zhi Li}.} \bibinfo{year}{2020}\natexlab{a}.
\newblock \showarticletitle{A TD3-based multi-agent deep reinforcement learning method in mixed cooperation-competition environment}.
\newblock \bibinfo{journal}{\emph{Neurocomputing}}  \bibinfo{volume}{411} (\bibinfo{year}{2020}), \bibinfo{pages}{206--215}.
\newblock
\showISSN{0925-2312}
\urldef\tempurl%
\url{https://doi.org/10.1016/j.neucom.2020.05.097}
\showDOI{\tempurl}


\bibitem[\protect\citeauthoryear{Zhang, Zhao, Wu, Li, Huang, and Tan}{Zhang et~al\mbox{.}}{2020b}]%
        {zhang2020cost}
\bibfield{author}{\bibinfo{person}{Yifan Zhang}, \bibinfo{person}{Peilin Zhao}, \bibinfo{person}{Qingyao Wu}, \bibinfo{person}{Bin Li}, \bibinfo{person}{Junzhou Huang}, {and} \bibinfo{person}{Mingkui Tan}.} \bibinfo{year}{2020}\natexlab{b}.
\newblock \showarticletitle{Cost-sensitive Portfolio Selection via Deep Reinforcement Learning}.
\newblock \bibinfo{journal}{\emph{IEEE Transactions on Knowledge and Data Engineering}} (\bibinfo{year}{2020}).
\newblock


\bibitem[\protect\citeauthoryear{Zheng, Zhu, Li, Chen, and Tan}{Zheng et~al\mbox{.}}{2019}]%
        {zheng2019finbrain}
\bibfield{author}{\bibinfo{person}{Xiao-lin Zheng}, \bibinfo{person}{Meng-ying Zhu}, \bibinfo{person}{Qi-bing Li}, \bibinfo{person}{Chao-chao Chen}, {and} \bibinfo{person}{Yan-chao Tan}.} \bibinfo{year}{2019}\natexlab{}.
\newblock \showarticletitle{FinBrain: when finance meets AI 2.0}.
\newblock \bibinfo{journal}{\emph{Frontiers of Information Technology \& Electronic Engineering}} \bibinfo{volume}{20}, \bibinfo{number}{7} (\bibinfo{year}{2019}), \bibinfo{pages}{914--924}.
\newblock


\end{thebibliography}


\end{document}